\documentclass[longbibliography]{appolb}
\usepackage{graphicx}
\usepackage{hyperref}
\usepackage{amsmath}
\received{January 2, 2026}

\begin{document}
\title{QUBITS AND VACUUM AMPLITUDES
\thanks{Presented at Matter To The Deepest Recent Developments In Physics Of Fundamental Interactions XLVI International Conference of Theoretical Physics, 14-19 September 2025, Katowice, Poland.}%
}
\author{Germ\'an Rodrigo
\address{Instituto de F\'{\i}sica Corpuscular, Universitat de Val\`{e}ncia -- Consejo Superior de Investigaciones Cient\'{\i}ficas, Parc Cient\'{\i}fic, E-46980 Paterna, Valencia, Spain.}
}
\maketitle
\begin{abstract}
High-energy colliders, such as the Large Hadron Collider (LHC) at CERN, are genuine quantum machines, so, in line with Richard Feynman’s original motivation for Quantum Computing, the scattering processes that take place there are natural candidates to be simulated on a quantum system. Potential applications range from quantum machine learning methods for collider data analysis, to faster and more precise evaluations of intricate multiloop Feynman diagrams, more efficient jet clustering, improved simulations of parton showers, and many other tasks. In this work, the focus will be on two specific applications: first, the identification of the causal structure of multiloop vacuum amplitudes, a key ingredient of the Loop–Tree Duality and an area with deep connections to graph theory; and second, the integration and sampling of high-dimensional functions. The latter constitutes a first step toward the realization of a fully fledged quantum event generator operating at high perturbative orders.
\end{abstract}
  
\section{Introduction}

High-energy physics is entering an era of unprecedented experimental precision. The upcoming High-Luminosity phase of the CERN's Large Hadron Collider (HL-LHC) will enable us to measure the fundamental properties of elementary particles, such as the couplings of the Higgs boson, with remarkable accuracy. This rapid advancement creates a strategic imperative for the theoretical physics community to advance theoretical predictions at a commensurate pace. The precision gap between theory and experiment is not a distant concern, but an imminent challenge. Projections~\cite{Cepeda:2019klc} from the ATLAS and CMS experiments for the HL-LHC suggest that, for several key measurements, the uncertainty associated with theoretical predictions will be a significant component of the total uncertainty, if not the dominant one. Regarding the establishment of any robust claim for potential discoveries, this scenario, in which our theoretical understanding is less precise than our experimental capability, is untenable and requires a paradigm shift in the computational techniques used to generate theoretical predictions.

The physical processes taking place in colliders such as the LHC are governed by Quantum Field Theory (QFT), since the interactions among particles are inherently Quantum Mechanical. This provides the basis for arguing that colliders are quantum machines and, in particular, that {\bf the LHC is the largest quantum machine ever constructed}. In this context, Quantum Computing (QC) emerges as the most natural framework for developing new algorithms and simulation tools for collider physics. Many collider-related computations are expected to be intrinsically well suited and potentially more efficient for quantum computers than for their classical counterparts. This perspective echoes Richard P. Feynman’s famous insight: {\it Nature isn’t classical, dammit, and if you want to make a simulation of nature, you better make it quantum}.

The core computational challenge in making theoretical predictions from QFT is the calculation of scattering amplitudes. These calculations are typically organized using Feynman diagrams, which provide a perturbative expansion representing all possible ways particles can interact. Each Feynman diagram or scattering amplitude corresponds to a complex, high-dimensional integral over the momenta of all possible intermediate or virtual particle states and the phase-space of the final-state particles. The precision required by the HL-LHC demands calculations involving diagrams with multiple loops and many external particles, thus exponentially increasing the complexity and dimensionality of these integrals.

\section{Qubits and Causality in Particle Physics}

A critical principle governing theoretical calculations is causality. A physical process must be causal, which means that the effects cannot precede their causes. Physically valid processes correspond to causal configurations in which no particle travels back in time. In the language of Feynman diagrams, this translates into a specific constraint on the flow of momentum or propagation path. That is, a particle cannot describe a closed cycle along a loop. In graph-theory terminology, these causal configurations are known as Directed Acyclic Graphs (DAGs). In contrast, noncausal, or cyclic, configurations are nonphysical because they imply a particle returning to its point of emission, and thus a violation of causality. 

The Feynman representation of scattering amplitudes does not require the explicit specification of particle propagation directions because their manifest Lorentz covariance allows us to encode all time-orderings simultaneously in a single compact expression. However, since this representation includes cyclic configurations, the integrand suffers from spurious singularities. A solution is provided by the Loop-Tree Duality (LTD), which is manifestly causal, i.e., it only encodes the acyclic or physical configurations, leading to integrands that are less singular and thus more suitable for numerical integration. A particularly interesting strategy is to base theoretical calculations on vacuum amplitudes~\cite{Ramirez-Uribe:2024rjg,LTD:2024yrb}, which provides additional conceptual and technical advantages.

At this point, we introduce a {\bf new sort of qubit}, the Feynman propagator. A Feynman propagator encodes the quantum superposition of a particle propagating between two interaction vertices in both directions, and can therefore be viewed as a two-state quantum system. In a quantum-mechanical notation, the following analogy can be introduced:
\begin{equation}
G_{\rm F} (q_i) = \frac{1}{q_i^2 - m_i^2 + \imath 0} \equiv \frac{1}{\sqrt{2}} \left( |0\rangle  + |1\rangle\right)~,
\end{equation}
where $|0\rangle$ denotes propagation in one direction and $|1\rangle$ in the opposite direction. A Feynman diagram, or scattering amplitude, corresponds to a quantum superposition of $2^n$ states, of which only a subset are acyclic and thus physically meaningful. Isolating these acyclic states, which encode the causal configurations, makes the role of causality in LTD manifest. Conversely, by fixing the propagation directions, one can bootstrap the corresponding integrand representation in LTD.

\section{Mapping Feynman Diagrams to Quantum Circuits}

The challenge of identifying the subset of causal DAGs  among all possible momentum-flow configurations in multiloop Feynman diagrams can be reframed as an unstructured search problem over an exponentially large space of configurations. This makes it a natural candidate for QC, specifically for search algorithms based on Grover’s principles~\cite{Grover:1997fa,Ramirez-Uribe:2021ubp}. Alternatively, the same problem can be formulated as a minimization task, where the goal is to find the ground state of a Hamiltonian whose energy counts the number of cycles in the graph. In this view, a Variational Quantum Eigensolver~(VQE) approach becomes applicable~\cite{Clemente:2022nll}.

Given the current limitations of quantum hardware, the primary goal is not necessarily to achieve a significant speedup in finding the physically relevant solutions, but rather to gain insight into how to design and optimize the quantum oracle. The general workflow of a Grover's based algorithm and the corresponding quantum circuit is outlined as follows. First, it requires a central set of qubits, the edge register, composed of a number of qubits equal to the number of internal Feynman propagators. These qubits are prepared in a uniform superposition of $|0\rangle$ and $|1\rangle$ states, typically achieved by applying Hadamard gates to each qubit. In this encoding, the state $|0\rangle$ for a given qubit represents the propagation of a particle between two interaction vertices in one predetermined direction, while the state $|1\rangle$ represents the propagation in the opposite direction. 

The oracle’s objective is to mark those entangled states in the edge register that correspond to valid DAG configurations, by applying a conditional phase flip without changing the underlying probability distribution of the qubits. This tagging may require additional ancillary qubits to store intermediate results or to perform arithmetic and logical checks on the edge assignments. After the oracle marks the causal states, a diffusion operator is applied. This operator performs a transformation in the Hilbert space that amplifies the probability amplitude of the marked states and suppresses that of the nonmarked states such that a measurement of the qubits in the edge register will select, with higher probability, a DAG state. Repetition of this procedure serves to identify all DAG configurations. 

An important innovation was introduced in Ref.~\cite{Ramirez-Uribe:2024wua} in the design of the oracle, which is built from multicontrolled Toffoli gates. Such a gate acts on a multitude of control qubits. If all the control qubits are in the same state, i.e. the $|1\rangle$ state, it flips the state of a target qubit by applying a $X$ gate, leaving the state of the target qubit unchanged otherwise. In other words, cycles in the Feynman diagram are equivalent to multicontrolled Toffoli gates, because the control qubits are all in the state $|1\rangle$ precisely when all associated propagators are oriented in the same direction. In addition to introducing another interesting entry in the dictionary between QC and Feynman diagrams, this oracle construction leads to a significant reduction in the implementation cost for certain classes of diagrams, greatly improving the practical run time on quantum simulators and potentially in quantum hardware.

\section{Optimizing the Oracle with Graph-Theory Principles}

\begin{figure}[t]
\begin{center}
\includegraphics[scale=.14]{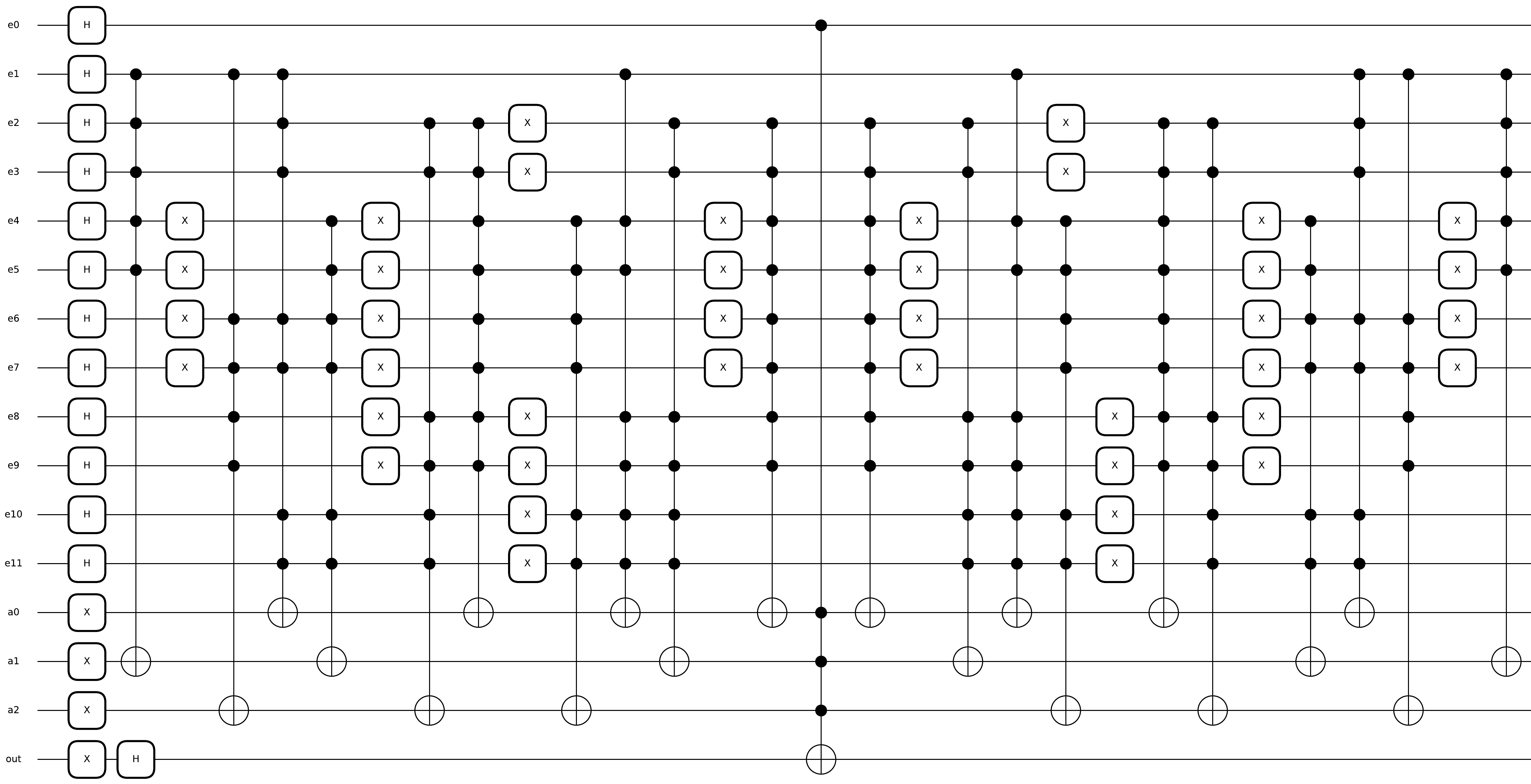}
\caption{Optimal oracle of a three-loop topology with $12$ propagators, represented by qubits $e_0$ to $e_{11}$. Only $3$ ancillary qubits ($a_0$ to $a_2$) are required.
\label{fig:oracle}}
\end{center}
\end{figure}

The efficiency of a Grover's based search depends critically on the resource requirements of its oracle. A significant advance comes from applying the principles of graph theory to optimize the oracle's design~\cite{Ochoa-Oregon:2025opz}. The key insight is to analyze the relationships between the different loop clauses, i.e. the conditions that define a cycle, and identify which loop clauses are mutually exclusive. A graph of Mutually Exclusive Clauses (MEC) is constructed and the problem of optimizing the oracle then becomes finding the Minimum Clique Partition (MCP), the smallest number of fully connected subgraphs (cliques) needed to cover all vertices in the MEC graph.

The impact of this optimization is a significant reduction in the number of ancillary qubits required. Since all clauses within a single clique are mutually exclusive, the information about them can be stored on a single ancillary qubit instead of one qubit per clause. For an illustrative three-loop example (Fig.~\ref{fig:oracle}), this technique reduces the required number of ancillary qubits from $7$ to just $3$. This reduction in ancillary qubit requirements may be the critical step to bring the analysis of physically interesting multiloop diagrams within the grasp of near-term, Noise Intermediate Scale Quantum~(NISQ) devices.

By combining the power of quantum search with sophisticated graph-theory optimization, it becomes possible to design efficient quantum oracles that require significantly fewer resources. This approach makes the identification of physically relevant configurations for high-order calculations a tractable problem, setting the stage for their subsequent integration.

\section{Quantum Integration of Scattering Amplitudes}

Once the causal configurations of a Feynman diagram or scattering amplitude have been determined, the next major computational hurdle is the evaluation of the associated  multidimensional integrals, with focus in LTD, where multiloop amplitudes are reformulated into well-behaved, manifestly causal integrand representations that are particularly well suited to both classical and quantum numerical techniques. Two novel quantum algorithms have been developed that specifically target this challenge, QFIAE~\cite{deLejarza:2023qxk} and QAIS~\cite{Pyretzidis:2025stx}. Other quantum integration methods include~\cite{Yi:2025feo,Williams:2025hza,Cruz-Martinez:2023vgs,Agliardi:2022ghn}.

\subsection{Quantum Fourier Iterative Amplitude Estimation (QFIAE)}

\begin{figure}[t]
\begin{center}
\begin{tabular}{c}
\includegraphics[scale=.20]{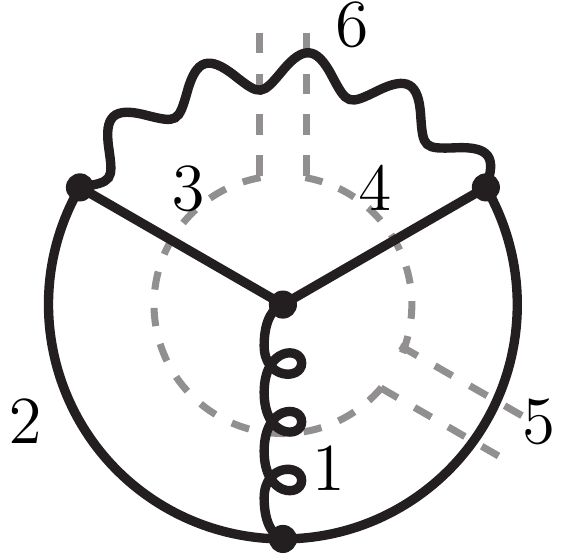} \\
\includegraphics[scale=.20]{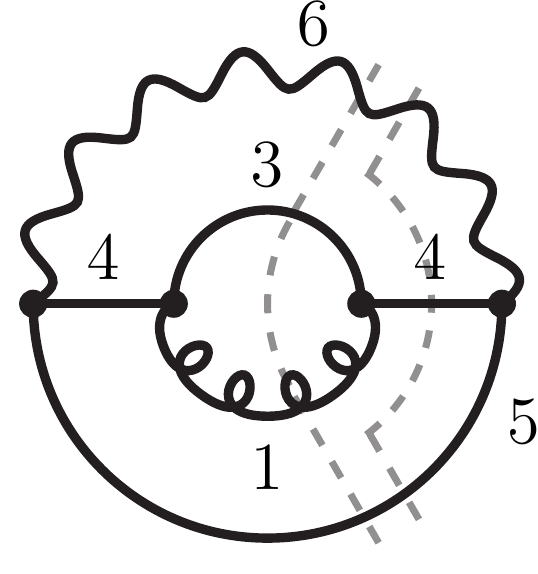} 
\end{tabular}
\begin{tabular}{c}
\includegraphics[scale=.44]{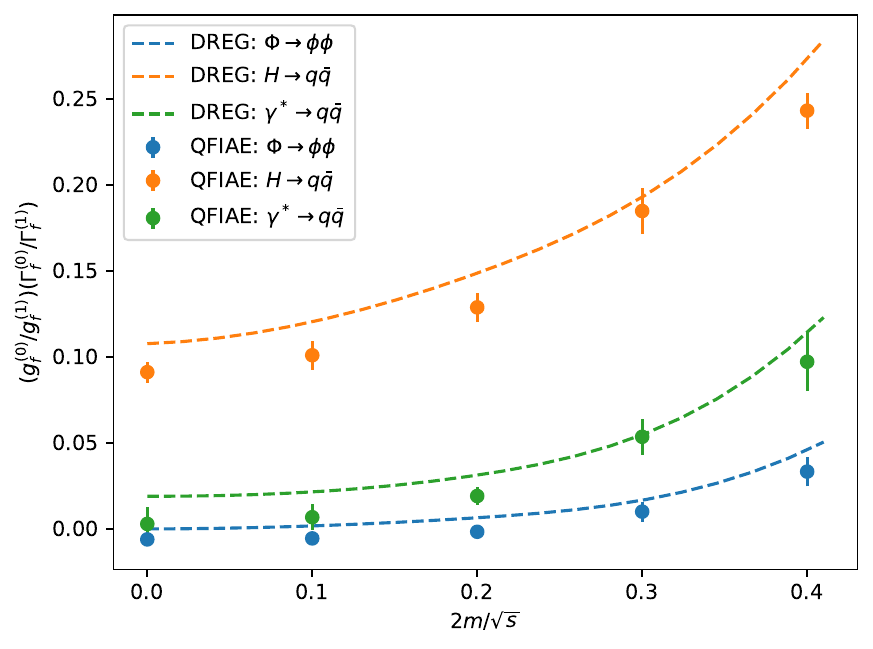}
\end{tabular}
\caption{Vacuum diagrams contributing to the decay rate $\gamma^*\to q\bar q (g)$ at NLO (left), and integrated results as a function of the quark mass using QFIAE partially in quantum hardware (right). 
\label{fig:lhc_integral}}
\end{center}
\end{figure}

The QFIAE~\cite{deLejarza:2023qxk} method is a hybrid approach that combines Quantum Machine Learning~(QML) with Quantum Amplitude Estimation~(QAE). The process involves two primary steps: 1) A Quantum Neural Network (QNN) is first trained to learn a compact and accurate Fourier series representation of the target integrand. This step effectively transforms a potentially complex function into a sum of simpler trigonometric components. 2) The Iterative Quantum Amplitude Estimation~(IQAE)~\cite{Grinko:2021iad} algorithm, a powerful variant of Grover's algorithm, used to integrate each trigonometric component of the learned Fourier series. The final integral is the sum of the individual results.

This method has been successfully applied to calculate Feynman integrals~\cite{deLejarza:2024pgk} and next-to-leading order (NLO) decay rates for several physical processes~\cite{deLejarza:2024scm}, using vacuum amplitudes in LTD as kernels for the loop and tree-level contributions. The close agreement between the results from quantum simulators and the established DREG method validates the QFIAE approach. While results in  hardware (Fig.~\ref{fig:lhc_integral}) show the effects of current device noise, they demonstrate the fundamental viability of the algorithm on real quantum processors.

\subsection{Quantum Adaptive Importance Sampling (QAIS)}

\begin{figure}[t]
\begin{center}
\includegraphics[scale=.28]{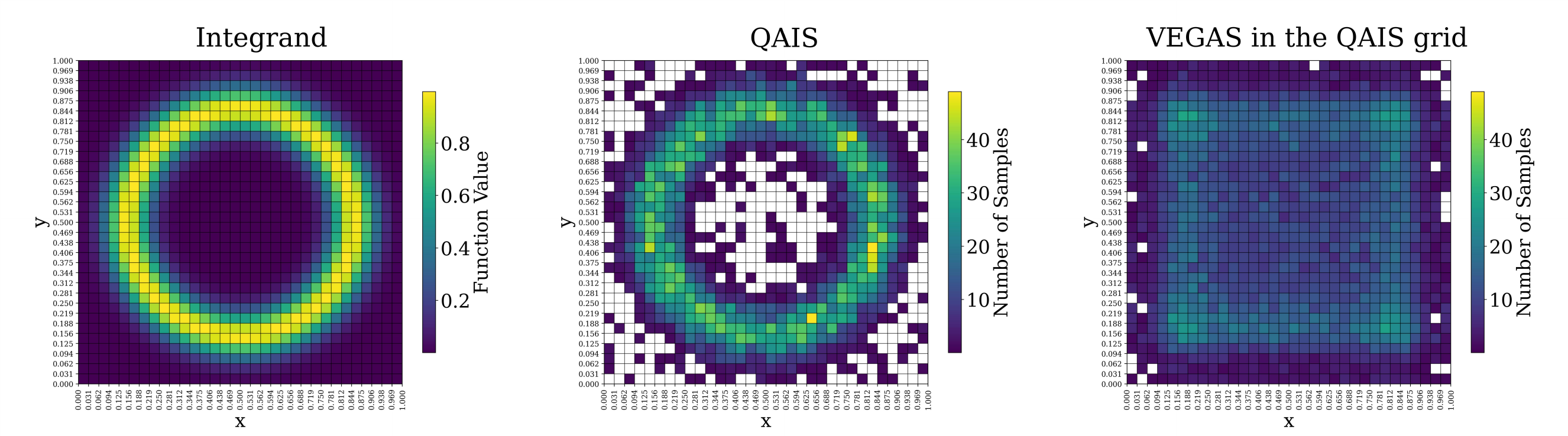}
\caption{Ring-shaped two-dimensional integrand and the corresponding PDFs in QAIS and VEGAS.
\label{fig:ring}}
\end{center}
\end{figure}

Classical Monte Carlo integration tools commonly employ Adaptive Importance Sampling methods to enhance efficiency, with VEGAS~\cite{Lepage:1977sw,Lepage:2020tgj,Hahn:2004fe} as the most prominent and widely used. However, these algorithms suffer from a crucial limitation. Because the complexity of the integration grid grows exponentially with the number of dimensions, they are constrained to use a separable grid, which performs poorly for integrands with intricate, correlated structures whose relevant regions are not aligned with the coordinate axes (Fig.~\ref{fig:ring}), often leading to phantom peaks, i.e. oversampled regions with little true contribution.

The Quantum Adaptive Importance Sampling (QAIS) algorithm~\cite{Pyretzidis:2025stx} is specifically designed to overcome this limitation. In this scenario, where complexity grows exponentially, a QC approach could provide a genuine quantum advantage, especially for integrals in very high dimensions. QAIS leverages a Parametrized Quantum Circuit (PQC) to define a non-separable proposal Probability Density Function (PDF) that is trained to approximate the target integrand, irrespective of its complexity. In this way, QAIS concentrates the PQC shots in the most relevant regions of the integration domain, maximizing sampling efficiency, substantially reducing the number of function evaluations required to attain a given precision, and naturally adapting to correlated structures that a separable grid is unable to capture. QAIS also incorporates a tiling component designed to correct the bias that mat arise when a finite number of shots causes the PQC to sample only a restricted portion of the full integration domain. This mechanism systematically covers the remaining regions of the grid, ensuring an unbiased estimator while retaining the sampling efficiency gains achieved by the learned PDF.

The good scaling of QAIS compared to VEGAS has been demonstrated, for example, by comparing their performance on a multipeak benchmark integral across increasing dimensions, and with a pentagon Feynman integral at one-loop, which is three-dimensional in LTD. QAIS consistently achieves a lower uncertainty for the same number of shots, with the performance gap widening as dimensionality increases.

\section{Conclusions}

High-energy colliders are genuine quantum machines, in which particle interactions are governed by the probabilistic and complex laws of QFT. This makes colliders a natural and compelling domain for the design and benchmarking of QC algorithms.
Causality is a fundamental principle in physics that, in the context of Feynman diagrams, determines the physically valid propagation directions of virtual particles. In graph-theory terms, physical configurations correspond to DAGs. A dictionary between particle physics and QC exists, where Feynman propapators are mapped to qubits, and cyclic configurations to multicontrolled Toffoli gates. This analogy enables the development of new QC algorithms and provides a framework to optimize quantum circuit design using concepts from graph theory.
We have developed and validated two quantum integration algorithms, QFIAE and QAIS, both of which provide viable and powerful tools for addressing the challenging multidimensional integrals that arise in high-energy physics. In particular, QAIS, with its capacity to capture highly correlated integrands and its favorable scaling with dimensionality, stands out as especially promising for the increasingly complex calculations required in the future.

\section*{acknowledgments}
This work is supported by the Spanish Government and ERDF/EU - Agencia Estatal de Investigaci\'on MCIN/AEI/10.13039/501100011033,  Grants No. PID2023-146220NB-I00, No. EUR2025-164820, and No. CEX2023-001292-S; and Generalitat Valenciana, Grant No. ASFAE/2022/009 (Planes Complementarios de I+D+i, NextGenerationEU).

\bibliographystyle{JHEP}
\bibliography{2025_MTTD_rodrigo}

\end{document}